\documentclass[conference]{IEEEtran}
\usepackage{epsfig}

\usepackage{enumitem}
\usepackage{amsmath}

\usepackage{balance}
\usepackage{xcolor, colortbl}
\usepackage{pifont}
\newcommand{\ignore}[1]{}

\title{\vspace{-0.1in} Touch\'e: Towards Ideal and Efficient Cache Compression By Mitigating Tag Area Overheads}

\begin{document}

\author{ \IEEEauthorblockN{Seokin Hong\IEEEauthorrefmark{1}, Bulent Abali\IEEEauthorrefmark{2}, Alper Buyuktosunoglu\IEEEauthorrefmark{2}, Michael B. Healy\IEEEauthorrefmark{2}, and Prashant J. Nair\IEEEauthorrefmark{3}}
    \IEEEauthorblockA{\\ \IEEEauthorrefmark{1}Kyungpook National University \hspace{0.2in} \IEEEauthorrefmark{2}IBM T. J. Watson Research Center \hspace{0.2in} \IEEEauthorrefmark{3}The University of British Columbia \\ \textit{seokin@knu.ac.kr}  \hspace{0.65in}  [\textit{abali,alperb,mbhealy}]\textit{@us.ibm.com}  \hspace{0.5in} \textit{prashantnair@ece.ubc.ca}}}
   
	\maketitle
	\begin{abstract}
	Compression is seen as a simple technique to increase the effective cache capacity. Unfortunately, compression techniques either incur tag area overheads or restrict data placement to only include neighboring compressed cache blocks to mitigate tag area overheads. Ideally, we should be able to place arbitrary compressed cache blocks without any placement restrictions and tag area overheads.
	
	This paper proposes Touch\'e, a framework that enables storing multiple arbitrary compressed cache blocks within a physical cacheline without any tag area overheads. The Touch\'e framework consists of three components. The first component, called the ``Signature'' (SIGN) engine, creates shortened signatures from the tag addresses of compressed blocks. Due to this, the SIGN engine can store multiple signatures in each tag entry. On a cache access, the physical cacheline is accessed only if there is a signature match (which has a negligible probability of false positive). The second component, called the ``Tag Appended Data'' (TADA) mechanism, stores the full tag addresses with data. TADA enables Touch\'e to detect false positive signature matches by ensuring that the actual tag address is available for comparison. The third component, called the ``Superblock Marker'' (SMARK) mechanism, uses a unique marker in the tag entry to indicate the occurrence of compressed cache blocks from neighboring physical addresses in the same cacheline. Touch\'e is completely hardware-based and achieves an average speedup of 12\% (ideal 13\%) when compared to an uncompressed baseline.
	
	\end{abstract}


\section{Introduction}
As Moore's Law slows down, the number of transistors-per-core for Last-Level caches (LLC) tends to be stagnating~\cite{mooreslaw1,mooreslaw2,mooreslaw3,dark1,dark2}. For instance, in moving from Ivy Bridge (i7-4930K processor at 22nm) to Broadwell (i5-5675C processor at 14nm), the LLC capacity per core (thread) has stagnated at 1MB~\cite{intelivy,intelbroadwell}. One can employ data compression to increase the effective LLC capacity~\cite{bdi,Arelakis:2014:SSC:2665671.2665696,ResidueCache,ZC}. Unfortunately, data compression may also incur significant tag area overheads~\cite{Alameldeen:2004,kim2002low,ECM}. This is because, in conventional caches each block needs a separate tag. We can reduce the tag area overheads by storing compressed blocks only from neighboring addresses~\cite{SCC,YACC,DCC1,DCC2}. This enables us to use a single overlapping tag for all compressed blocks. However, such an approach restricts data compression only to regions that contain contiguous compressed blocks. Ideally, we would like to employ LLC compression without any data placement restrictions and tag area overheads.

Tag overheads are a key roadblock for cache compression. For instance, if we store 4x more blocks, the effective LLC capacity can be increased by 4x. But we will also incur the area overheads for maintaining 4x unique tags. Furthermore, it is likely that these unique tags have no locality, cannot be combined together, and therefore incur significant area overheads~\cite{DCC1,DCC2}. One can reduce the tag area overhead with placement restrictions. For instance, if we set a rule that only compressed blocks from neighboring memory addresses can reside in a physical cacheline, then we can overlap their tags. These contiguous compressed blocks are called ``superblock'' and their tag is called a ``superblock-tag''~\cite{SCC, YACC}. For a 4MB 8-way cache, superblock-tags can track 4 compressed blocks per cacheline with 1.35x tag area. 

\begin{figure}[h!]
  \vspace{-0.12in}
  \centering \centerline{\psfig{file=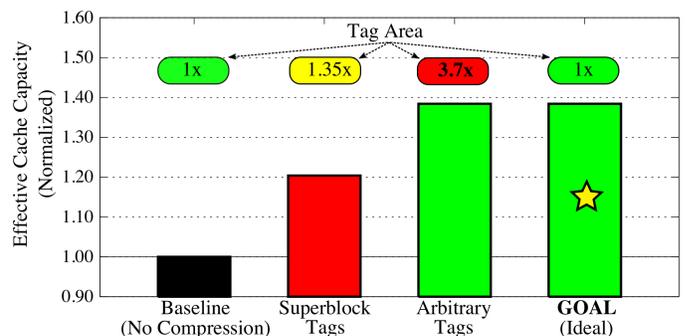,width=\columnwidth}}
  \vspace{-0.12in}
  \caption{The effective capacity and tag area overheads for a 4MB last-level cache employing compression. Superblock-tags uses 1.35x tag area while providing 20\% higher effective capacity. Arbitrary-tags uses 3.7x tag area while providing 38\% higher effective capacity. The goal of this paper is to obtain 38\% higher effective capacity with no tag overhead.}
  \label{fig:figure1}
  \vspace{-0.1in}
\end{figure}

Restricting block placement by using superblocks reduces the benefits of compression. Figure~\ref{fig:figure1} shows the effective LLC capacity for four designs executing 29 memory-intensive SPEC workloads in mixed and rate modes on a 4MB shared LLC~\cite{spec}. The first design is a baseline LLC without data compression. The second design employs data compression in LLC while using superblocks. While such a design has a tag area of 1.35x as compared to the baseline, it also increases the effective LLC capacity only by 20\%. This is because only blocks from neighboring addresses can be compressed and stored in the cacheline. The third design enables data compression to place arbitrary tags in the same cacheline. While this design increases the effective LLC capacity by 38\%, it also requires 3.7x the tag area. The fourth design is an ideal design which places arbitrary tags in the same cacheline without any area overheads. This paper presents Touch\'e, a framework that helps achieve the fourth design to enable near-ideal LLC compression.

Touch\'e mitigates the tag area overheads by using a shortened signature of the full tag address for each compressed block. This has two key benefits. First, short signatures require fewer bits as compared to full tag addresses. Due to this, multiple signatures from different tags addresses can be placed in the space that was originally reserved for only a single tag address. Second, by enabling arbitrary signatures to reside next to each other, we can overcome restrictions of prior work that require compressed blocks to be from neighboring addresses. Furthermore, as compression creates unused space in the data array, tag addresses can be appended to compressed blocks and stored in this unused space. 

The Touch\'e framework consists of three components. The first component, called the ``Signature'' (SIGN) engine, creates shortened signatures from the tag addresses and places them in the tag array. The second component, called the ``Tag Appended Data'' (TADA) mechanism, appends full tag addresses to the compressed blocks and stores them in the data array. The third component, called the `` Superblock Marker'' (SMARK) mechanism, uses a unique marker in the tag-bits to enable Touch\'e to identify superblocks that contain 4 contiguous compressed blocks from neighboring physical addresses. We describe each mechanism below:

\begin{enumerate}[leftmargin=0cm,itemindent=.5cm,labelwidth=\itemindent,labelsep=0cm,align=left, listparindent=0.4cm]
\item \textbf{Signature (SIGN) Engine}: The Touch\'e framework is implemented within the LLC controller. The core provides the LLC controller with a 48-bit physical address for each request~\footnote{Processor vendors have already proposed schemes like the Intel 5-level paging for enabling 57-bit physical addresses to increase the physical address space from 256 TB to 128 PB~\cite{intel5level}. This would increase the tag address bits within an LLC tag entry by 9.}. The LLC controller uses this physical address to index into the appropriate set. At the same time, Touch\'e invokes the SIGN engine to create a shortened 9-bit signature of the tag and looks up all the ways for a matching signature. On a signature match, the corresponding compressed block is accessed from the data array. As these signatures are only 9-bits long, several signatures, each belonging to a different tag address, can co-reside in a tag entry. For instance, a 4MB 8-way LLC with 64 Byte cachelines has tag entries that store 29-bit tag address. The SIGN engine can store up to three 9-bit signatures in the space that was designed for a single 29-bit tag address. This enables Touch\'e to store up to three arbitrary compressed blocks without any tag area overheads.

Unfortunately, simply using shortened 9-bit signatures can lead to false positive tag matches (signature collisions). Signature collisions cause the LLC controller to incorrectly access blocks that do not have matching tag addresses for each access. For instance, in a workload that has a 0\% cache hit-rate (worst case scenario), a 9-bit signature has an average signature collision rate of 0.19\% (i.e., $\frac{1}{2^{9}}$). Furthermore, as each way in a set can have up to three 9-bit signatures, Touch\'e potentially needs to check \textit{twenty-four} 9-bit signatures in an 8-way LLC (worst case scenario) which results in a signature-collision rate of 4.58\%. Therefore, it is essential to also check the full tag address on a signature collision.

\item \textbf{Tag Appended Data (TADA) Mechanism}: The full tag addresses of the compressed blocks can be stored in the data array. Touch\'e re-provisions a portion of the additional space that is obtained by compression to store the full tag addresses. To this end, Touch\'e uses the ``Tag Appended Data'' (TADA) mechanism to append full tags beside the compressed blocks. The TADA mechanism appends metadata information on compressibility (3-bits), dirty and valid state (2-bits), and the full tag address (29-bits) to each compressed block. Overall, TADA increases the block size by only 34 bits (4.25 Bytes) and our experiments show that it only minimally reduces the effective LLC capacity. On an access, TADA interprets the last few bits in a compressed cacheline as metadata and tag addresses. As TADA checks the full tags on all signature matches and collisions, it guarantees the correctness of each LLC access.

\item \textbf{Superblock Marker (SMARK) Mechanism}: Shortened 9-bit signatures enable storing up to three compressed blocks. However, there can be instances of four compressed blocks from neighboring addresses (superblock). To address this scenario, Touch\'e uses a ``Superblock Marker'' (SMARK) mechanism to generate a unique 16-bit marker. Touch\'e stores this 16-bit marker in the tag bits, and uses this marker to indicate the presence of a superblock within the cacheline. 

With a negligible probability (0.012\%), the unique 16-bit marker can flag a match with the signatures that are stored by the SIGN engine. We call these scenarios as SMARK collisions. Fortunately, SMARK collisions cause no correctness problems. This is because even after a marker collision, the TADA mechanism will read the data array and check for full tag matches. During a collision, the tag addresses will not match and the compressed blocks are not processed by the LLC. The SMARK mechanism enables Touch\'e to derive all the benefits of superblocks while also enabling the storage of up to three arbitrary compressed blocks.
\end{enumerate}

Touch\'e provides a speedup of 12\% (ideal 13\%) without any area overheads. Touch\'e requires comparators and lookup tables within the LLC controller. Touch\'e is a completely hardware-based framework that enables near-ideal compression.

	\section{Background and Motivation}
We provide a brief background on last-level cache organization and highlight the potential of data compression.

\subsection{Last-Level Caches: Why Size Matters}
Processors tend to have several levels of on-chip caches. Caches are designed to exploit spatial and temporal locality of accesses. Due to this, caches help improve the performance of processors as they reduce the number of off-chip accesses and reduce the latency of memory requests. Caches are usually designed such that each level is progressively larger than its previous level. Consequently, the Last-Level Cache (LLC) tends to have the largest size, is typically shared, and occupies significant on-chip real-estate. Due to this, it is beneficial to increase the LLC capacity per core as this would enable the designers to fit a larger number of blocks on-chip and further reduce the number of off-chip accesses~\cite{isscc:sram}.

\subsection{Last-Level Caches: Capacity Stagnation}
\begin{figure}[htb!]
\vspace{-0.15in}
  \centering \centerline{\psfig{file=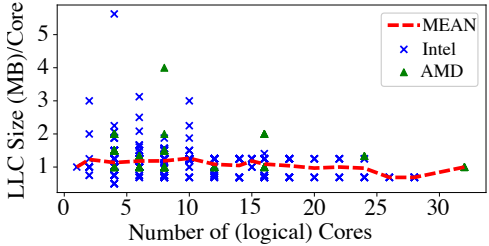,width=0.75\columnwidth}}
  \caption{The Last-Level Cache (LLC) capacity per (logical) core for Intel and AMD processors from 2009 to 2018. On average, as the number of cores has increased, the LLC capacity per core has reduced.}
  \label{fig:trendcommercial}
 \vspace{-0.15in}
\end{figure}

Figure~\ref{fig:trendcommercial} shows the LLC capacity per core for commercial Intel and AMD processors from 2009 until 2018. On average, as the number of cores has increased, the LLC capacity per core has reduced. In current multi-core systems, the LLC capacity per core tends to be less than 1MB. Therefore, going into the future, it is beneficial to look at techniques to improve the effective capacity of the LLC~\cite{sudoku}.


\subsection{Last-Level Caches: Organization}
\begin{figure}[h!]
\vspace{-0.1in}
  \centering \centerline{\psfig{file=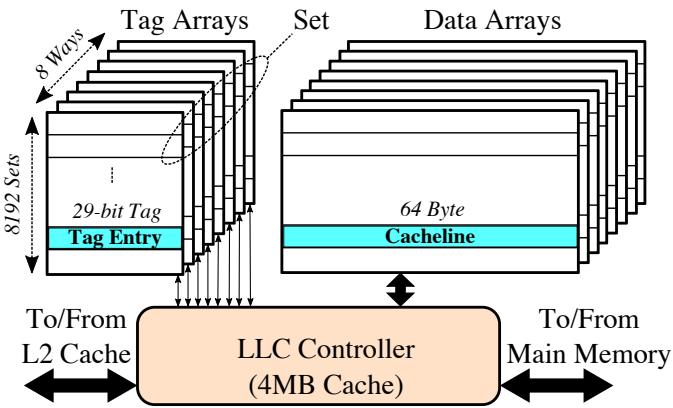,width=\columnwidth}}
  \caption{The organization of a 4MB Last-Level Cache (LLC). The LLC consists of data arrays, tag arrays, and an LLC controller. The tags are 29-bits long and all tag entries across the ways in a set are searched in parallel.}
  \label{fig:LLCorg}
 \vspace{-0.1in}
\end{figure}

A Last-Level Cache (LLC) is organized into data arrays and tag arrays. Each cacheline in the data array has a corresponding tag entry in the tag array. Furthermore, groups of cachelines form ``sets'' and each cacheline in a set corresponds to a separate ``way''. As the size of the LLC is significantly smaller than the total physical address space, several blocks can map into the same set. Because of this, the LLC controller stores a tag address in the tag entry to uniquely identify the block in the cacheline. For instance, as shown in Figure~\ref{fig:LLCorg}, a 4MB 8-way LLC with 64-byte lines, uses 29 bits of tag address. On a cache access, all the tag entries for each of the ways in a set is searched in parallel by the LLC controller.

\subsection{Compression: Higher Effective Capacity}
Several prior works have proposed using efficient and low-latency algorithms to compress blocks, thereby storing more blocks and improving the effective LLC capacity. Typically, LLC compression techniques are typically implemented within the LLC controller.
\subsubsection{Efficient Data Compression Algorithms}
The Base Delta Immediate (BDI) and Frequent Pattern Compression (FPC) are two state-of-the-art low-latency compression algorithms~\cite{bdi,fpc}. BDI uses the insight that data values tend to be similar within a block and therefore can be compressed by representing them using small offsets. FPC uses the insight that blocks contain frequent patterns like all-zeros, all-ones, etc. FPC represents frequent patterns with fewer bits. Prior work has shown that both BDI and FPC can be implemented to execute with a a single-cycle delay and can be implemented within the LLC controller~\cite{bdi,fpc}. 

\begin{figure}[h!]
  \centering \centerline{\psfig{file=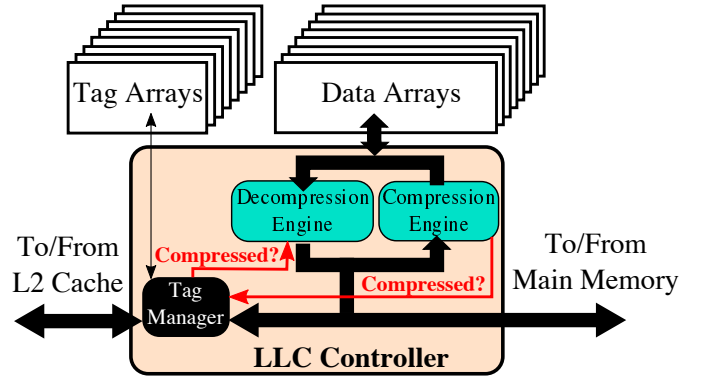,width=\columnwidth}}
  \caption{The LLC compression-decompression engine. The compression-decompression engine taps the data bus and stores compressibility information in the tag entries.}
  \label{fig:LLCcompression}
\end{figure}

\subsubsection{Compression-Decompression Engine}
As shown in Figure~\ref{fig:LLCcompression}, the LLC controller implements a compression-decompression engine that taps the bus going into the cache data array. The LLC controller contains a separate ``tag manager'' to manage tag entries. The compression-decompression engine implements both BDI and FPC and chooses the best algorithm. The tag manager maintains the compressibility information in the tag entries.

\subsubsection{Distribution of Compressed Data Size}  
\begin{figure}[htb!]
 \vspace{-0.1in}
  \centering
  \centerline{\psfig{file=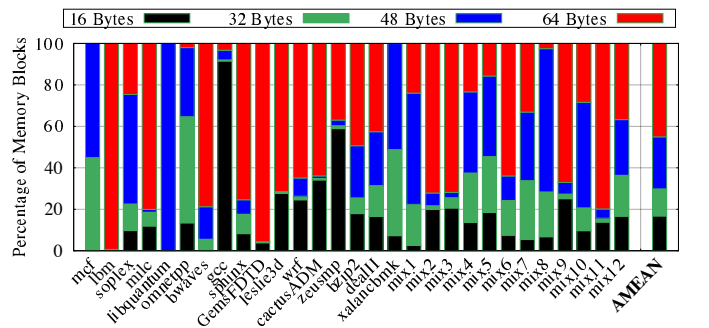,width=\columnwidth}}
   \vspace{-0.1in}
  \caption{The distribution of block-size for 29 SPEC workloads (rate and mix modes). On average, up to 55\% of the blocks can be compressed to 48 Bytes.}
  \label{fig:DataSize}
 \vspace{-0.2in}
\end{figure}

Figure~\ref{fig:DataSize} shows the distribution of the size of blocks after compression for 29 SPEC workloads. On average, 55\% of the blocks can be compressed to less than 48 bytes in size. Furthermore, 17\% of the lines can be compressed to be less than 16 bytes in size. Therefore, several workloads tend to have blocks with low entropy and can benefit from compression.
\newpage
\subsection{LLC Compression: Tag Area Overheads}
\begin{figure}[h!]
  \vspace{-0.1in}
  \centering \centerline{\psfig{file=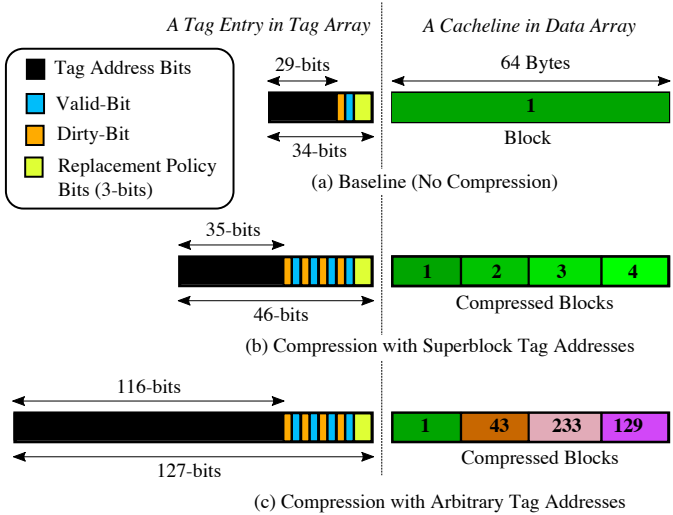,width=\columnwidth}}
      \vspace{-0.1in}
  \caption{The tag area overheads for different techniques. (a) The baseline technique that does not employ any compression has no tag area overheads. (b) The superblock technique increases the tag area to 1.35x. (c) Storing arbitrary tags increases the tag area to 3.7x. }
  \label{fig:tagoverheads}
  \vspace{-0.1in}
\end{figure}

Modern computing systems tend to operate on 64-byte blocks. Figure~\ref{fig:tagoverheads} (a) shows the design of the tag entry and the cacheline in the data array for a 4MB 8-way LLC that does not employ compression. The tag entry for each block requires a valid bit and a dirty bit. Furthermore, we assume that replacement policy is maintained at the cacheline-level and the largest block in the selected cacheline is evicted. To reduce the number of encoding bits in the tag array, blocks are compressed into 16 byte, 32 byte, or 64 byte boundaries. To reduce the number of bits in the tag entry further, one can restrict cachelines to store blocks only from contiguous addresses. Such a contiguous set of blocks is called a superblock. Prior work has shown that superblocks with 4 compressed blocks can reduce the tag area overheads to 8 bits. As shown in Figure~\ref{fig:tagoverheads} (b), a 4MB 8-way LLC that stores up to four blocks per cacheline will require 46 bits of tag entry. While superblocks help reduce tag area overheads, they limit the potential benefits of LLC compression as they restrict block placement to include only neighboring addresses. If one can store blocks from arbitrary addresses, we can unlock all the benefits of LLC compression. However, the disadvantage of this approach is that, as shown in Figure~\ref{fig:tagoverheads} (c), a 4MB 8-way LLC that stores up to four blocks per cacheline will require 127 bits of tag entry (3.7x higher than the baseline). 

\subsection{LLC Compression: Potential}
\begin{figure}[t!]
  \vspace{-0.1in}
  \centering \centerline{\psfig{file=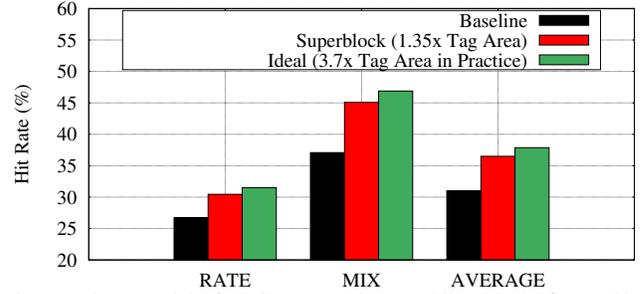,width=\columnwidth}}
  \vspace{-0.2in}
  \caption{The potential of LLC compression. Enabling blocks from arbitrary addresses increases the average hit-rate of the LLC from 31.5\% to 38.5\%.}
  \label{fig:Potential}
    \vspace{-0.15in}
\end{figure}

Figure~\ref{fig:Potential} shows the overheads and benefits of LLC compression for three techniques. The baseline technique does not employ compression, has no tag overheads and has an average hit-rate of 31.5\%. The second technique employs superblocks for compression, has a tag area of 1.35x and increases the average hit-rate to 36\%. The third technique highlights the potential hit-rate with compression when each cacheline can store up to 4 compressed blocks. Unfortunately, the third technique uses a tag area of 3.7x while also increasing the average LLC hit-rate to 38.5\%. 



	\section{The Touch\'e Framework}
\subsection{An Overview}
\begin{figure}[h!]
    \vspace{-0.05in}
  \centering \centerline{\psfig{file=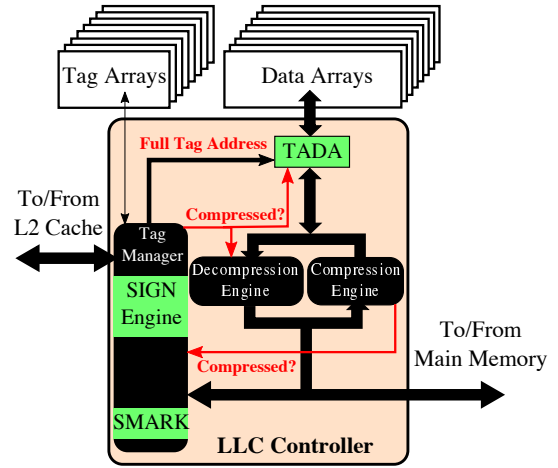,width=0.8\columnwidth}}
     \vspace{-0.05in}
  \caption{An overview of Touch\'e. Touch\'e consists of three components. The Signature (SIGN) Engine, the Tag Appended Data (TADA) mechanism, and the Superblock Marker (SMARK) mechanism. All components are implemented in the LLC controller with no changes to the LLC.}
  \label{fig:ToucheOverview}
    \vspace{-0.15in}
\end{figure}

Figure~\ref{fig:ToucheOverview} shows an overview of the Touch\'e framework. Touch\'e consists of three components. The first component, called the Signature (SIGN) Engine, generates shortened signatures of the tag addresses. The SIGN engine is designed within the tag manager. The second component, called the Tag Appended Data (TADA) mechanism, attaches full tag addresses to compressed memory blocks. The TADA mechanism taps the data bus after the compression-decompression engine and obtains the full tag address from the tag manager. The third component, called Superblock Marker (SMARK) mechanism, keeps track of superblocks by using a unique 16-bit marker in the tag entry. The SMARK mechanism is implemented in the tag manager. Touch\'e requires changes only in the LLC controller.

\subsection{Signature (SIGN) Engine}
The Signature (SIGN) Engine is implemented in the tag manager. The SIGN Engine generates shortened signatures from the full tag addresses supplied during the read and write accesses to the LLC.
\subsubsection{Identifying Compressed blocks}
On a LLC write, the compression-decompression engine informs the tag manager if the block is compressible; a block can be compressed to 16B, 32B or 48B. The tag manager uses the original valid bit and the dirty bit in its tag entry to encode this information. We use the insight that, for uncompressed blocks, the valid bit and the dirty bit can only exist in three states. For instance, a cacheline cannot be marked both invalid and dirty at the same time. The tag manager uses this unused state to flag cachelines that contains compressed blocks. Thereafter, for a cacheline that stores compressed blocks, the 1$^{st}$ and 2$^{nd}$ bits of the tag address encodes its valid bit and dirty bit. 

As shown in Table~\ref{table:compressedline}, on a read, the tag manager checks the original dirty bit and the valid bit in the tag entry to identify if the cacheline contains compressed blocks. If the cacheline is deemed to contain compressed blocks, the tag manager reads the 1$^{st}$ and 2$^{nd}$ bits from the tag address to determine if any of the cacheline contains blocks that are valid, dirty or both.
\begin {table}[h!]
\vspace{-0.05in}
\begin{center} 
\caption{Identifying Compressed blocks}{
\vspace{-0.15in}
\resizebox{\columnwidth}{!}{
\begin{tabular}{| l | c | c | c | c | }
\hline
                 Cacheline Status               &  Valid    & Dirty     &  Tag Address     & Tag Address      \\
                                                &  Bit      &  Bit      &  1$^{st}$ Bit  & 2$^{nd}$ Bit   \\ \hline \hline
                 Invalid            &   0           &  0    &   N/A  &  N/A       \\ \hline
                 Uncompressed: Valid            &   1           &  0    &   N/A  &  N/A       \\ \hline
                 Uncompressed: Valid and Dirty  &   1           &  1    &   N/A  &  N/A         \\ \hline \hline
                 \textbf{Compressed: Valid}     &   \textbf{0}  &  \textbf{1}  &   \textbf{1} &  \textbf{0}     \\ \hline
                 \textbf{Compressed: Valid and Dirty}     &   \textbf{0}  &  \textbf{1}  &   \textbf{1}  & \textbf{1}     \\ \hline
\end{tabular}}
}
\label{table:compressedline}
\end{center}
\vspace{-0.15in}
\end{table}

\subsubsection{Using Shortened Signatures}
To store multiple signatures within a single tag entry, the SIGN engine shortens the full tag address into a 9-bit signature. For a 4MB 8-way LLC, the full tag address is 29-bits long. For a compressed block, as the top 2 bits of the tag address space in its tag entry are already used for valid and dirty bits, we have 27 unused bits remaining in the tag address space of its tag entry. Therefore, we can store up to three 9-bit signatures corresponding to three compressed blocks. 

\begin{figure}[h!]
  \centering \centerline{\psfig{file=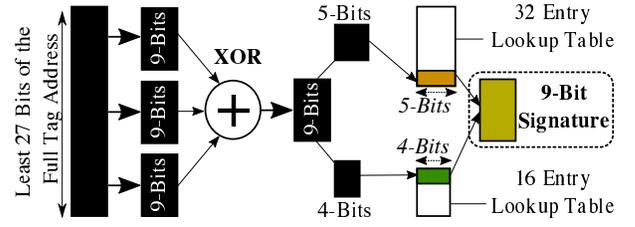,width=0.9\columnwidth}}
  \vspace{-0.1in}
  \caption{The signature generator within the SIGN Engine. The signature generator only requires one XOR operation and two parallel table lookups for each LLC access.}
  \label{fig:SIGNGen}
  \vspace{-0.1in}
\end{figure}

Figure~\ref{fig:SIGNGen} shows the design of the signature generator in the SIGN engine. The signature generator uses the least 27-bits of the full tag address and divides it into three 9-bit segments. Each bit of these 9-bit segments is then XORed together to generate a 9-bit output. The 9-bit output is then partitioned into a 4-bit segment containing its lowest bits and a 5-bit segment that contain its highest bits. These 4-bit and 5-bit partitions then index into a 16 entry lookup table and a 32 entry lookup table respectively. Each entry in lookup tables are populated at boot-time with unique numbers. The indexed 4-bit and 5-bit numbers from the lookup tables are then appended together to form a 9-bit signature. The overall latency of generating signatures is the delay of one 3-bit XOR gate and one parallel table lookup. For a high-performance processor executing at 3.2GHz, we estimate the signature generation to incur only 1 cycle. Furthermore, the latency of signature generation is masked by the latency of reading the tag entries for each LLC access (up to 5 cycles).

\subsubsection{Checking for Matching Signatures}
On an LLC read, the tag manager reads the tag entries from all the ways of the indexed set. At the same time, the SIGN engine forwards its 9-bit signature to the tag manager. The tag manager identifies if the cacheline contains compressed blocks using the original valid and dirty bits. For an uncompressed cacheline, the tag manager ignores the signature and uses the full tag address to check for a match.

If the cacheline contains compressed blocks, the tag manager ignores the first two bits of the tag address as they are valid and dirty bits. Thereafter, the remaining 27 bits in the tag address space of the tag entry are partitioned into three 9-bit entries. The tag manager then compares each of these three 9-bit entries with the 9-bit signature from the SIGN Engine. If the 9-bit entry does not match the 9-bit signature, then the block is guaranteed to be absent. On the other hand, if the 9-bit signature matches in any one of the ways, then the block is likely to be present. As a 9-bit signature is smaller than its full 29-bit tag address, there is a small chance of 0.19\% ($\frac{1}{512}$) that each 9-bit entry comparison with the 9-bit signature can result in a false positive match. We call these false positive matches of signatures as ``signature collisions''.

\subsubsection{Collision Rate of Signatures}
\begin{figure}[htb!]
  \vspace{-0.1in}
  \centering \centerline{\psfig{file=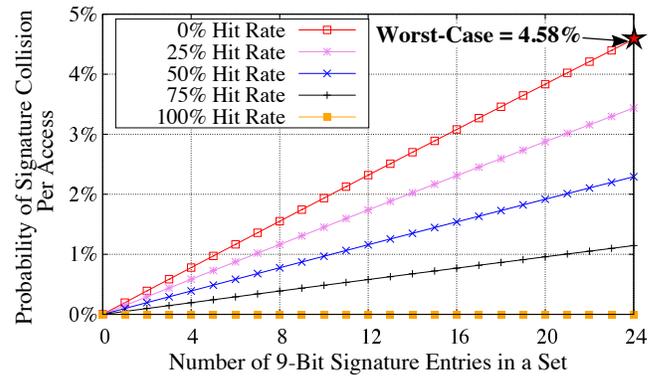,width=0.95\columnwidth}}
      \vspace{-0.1in}
  \caption{The probability of collision for a 9-bit signature as the number of signature entries vary in a set. In the worst-case, for a 8-way LLC, we expect a signature collision 4.58\% of the time for each access.}
  \label{fig:collision-rate}
    \vspace{-0.15in}
\end{figure}

As each tag entry can store up to three 9-bit signatures, an 8-way LLC would require up to twenty-four 9-bit signature comparisons for each access. As signatures are shorter than full tags, several tags may map into the same signature. As we perform twenty-four signature checks (in the worst-case), it is likely that some of LLC accesses will result in signature collisions. Figure~\ref{fig:collision-rate} shows the probability of collisions as the number of signatures present in the 8-ways varies from zero (all ways are uncompressed) to twenty-four (all ways have three compressed blocks) for different LLC hit-rates. 

In the worst case, we can expect a collision 4.58\% of the time and this occurs for a workload that has 0\% hit-rate. As signature collisions can cause the LLC to forward blocks with incorrect tag addresses to the processing cores, it is essential to check full tags.

\subsection{Tag Appended Data (TADA) Mechanism}
The Tag Appended Data (TADA) mechanism is implemented in the LLC controller and taps the data-bus between the compression-decompression engine and the data array.
\subsubsection{Appending Full Tag Addresses to Data}
\begin{figure}[h!]
      \vspace{-0.1in}
  \centering \centerline{\psfig{file=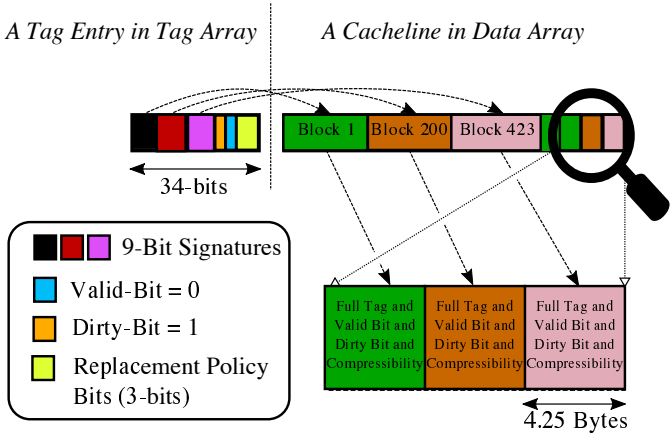,width=\columnwidth}}
      \vspace{-0.1in}
  \caption{A cacheline storing compressed blocks with TADA mechanism. The TADA mechanism appends full tag addresses, valid bit, dirty bit, and compressibility information for each block at the end of the cacheline.}
  \label{fig:tada}
    \vspace{-0.1in}

\end{figure}

During an LLC write, the TADA mechanism uses the full tags that are supplied by the tag manager. The TADA mechanism then appends the full tag addresses (29 bits), the valid-bit, the dirty-bit, and the compressibility information (3 bits) to the end of the cacheline (total 34 bits or 4.25 Bytes). Figure~\ref{fig:tada} shows a cacheline storing three compressed blocks and the TADA mechanism appending the information for each of these blocks at the end of the cacheline.

\subsubsection{Appending Full Tag Addresses to Data}
\begin{figure}[h!]
  \vspace{-0.1in}
  \centering \centerline{\psfig{file=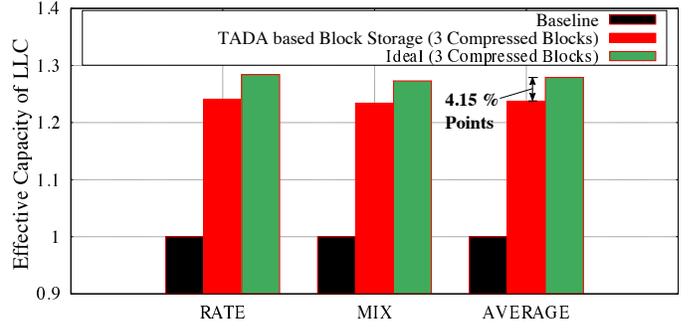,width=\columnwidth}}
  \vspace{-0.1in}
  \caption{The reduction in the effective LLC capacity by the TADA storage overhead. While TADA uses 4.25 Bytes per compressed block, it decreases the effective LLC capacity only by 4.15\% points as compared to an ideal scheme that does not require the metadata storage in the data array.}
  \label{fig:tadaeval}
    \vspace{-0.1in}
\end{figure}

The TADA mechanism appends 34 bits (4.25 Bytes) of information to the end of the cacheline containing compressed blocks. As a result, TADA reduces the space available to store the compressed block. We can store three 16B compressed blocks or a pair of a 32B and a 16B compressed blocks in the data array; the block size is stored in the compressibility information field. Fortunately, this additional loss of space only causes a few lines to reduce their effective capacity. Figure~\ref{fig:tadaeval} shows the reduction in effective LLC capacity due to TADA for an LLC that can store up to 3 arbitrary compressed blocks. TADA decreases the effective cache capacity by only 4.15\% points as compared to an ideal scheme that can store three arbitrary compressed blocks without any storage overheads.

\subsubsection{Detecting Collisions of Signatures}
TADA helps detect signature collisions. This is because, on a signature collision, the cachelines from the selected way(s) in the data array are read by the tag manager. The TADA mechanism extracts the full tag address from the cachelines and checks if they match the full tag address of the LLC access. If there is no match, TADA flags a signature collision. Therefore, TADA guarantees the detection of signature collisions and thereby ensures correctness. Furthermore, TADA extracts the the compressibility information and supplies to the decompression engine. The valid and dirty bits of compressed blocks are also stored using TADA. Therefore, TADA helps to avoid using any additional bits in the tag entry to store additional information.

\subsection{Latency Overheads} \label{Sig_coll_lat_overheads}
As the data array needs to be accessed during a signature collision, it can increase the LLC access latency.
\subsubsection{Additional Accesses to Data Arrays}
In the baseline system, an LLC access probes all the ways in the indexed set from the tag array. The cacheline from the data array is read-only in case of an LLC hit. As a tag access occurs for every access, irrespective of whether the access is a hit or a miss, the tag array is designed with lower access latency as compared to the data array. Typically, accessing the LLC tag array incurs a latency overhead of only 5 cycles. On the other hand, reading the LLC data array incurs an overhead of 30 cycles in modern processors~\cite{intelllclatency}.

In the Touch\'e framework, the data arrays are likely to be accessed even in case of an LLC miss. This is because the SIGN engine may incur signature collisions and may invoke the TADA mechanism to access the data array to detect signature collisions. In the worst-case, for a workload with 0\% hit-rate, this scenario may occur only 4.58\% of the times. Therefore, signature collisions will increase the overall latency of LLC access. Table~\ref{table:collisionData} shows the average latency of an LLC access during a collision.

\begin {table}[h!]
\vspace{-0.05in}
\begin{center} 
\caption{Additional Data Arrays Accessed on a Collision}{
\vspace{-0.15in}
\resizebox{0.95\columnwidth}{!}{
\begin{tabular}{| c | c | c |}
\hline
                 Number of Data Arrays Accessed     &  Probability & Latency (cycles)     \\ \hline \hline
                 1          &   0.9768           &  35        \\ \hline
                 2          &   0.0229           &  70            \\ \hline
                 3+          &   0.0003          &  105+         \\ \hline \hline
                 \textbf{Average: 1.0235}   &   1        &  \textbf{35.82}        \\ \hline
\end{tabular}}
}
\label{table:collisionData}
\end{center}
\vspace{-0.1in}
\end{table}

In the worst-case, all accesses can be a cache miss and a collision can occur 4.58\% of the times. As shown in Table~\ref{table:collisionData}, collisions increase the access latency to 35.82 cycles. For a worst-case workload with a 0\% hit-rate, the increase in the LLC tag access latency is denoted by Equation~\ref{eqn:collisionlatency}.
\begin{equation}\label{eqn:collisionlatency}
\normalsize{
\begin{aligned}
\text{New Tag Access Latency} = (1-0.0458)\times \text{Old Latency} + \\ 0.0458\times\text{Collision Latency}
\end{aligned}
}
\end{equation}

As the old tag access latency is 5 cycles and the collision latency is 35.82 cycles, the new tag access latency of Touch\'e is 6.4 cycles.

\subsubsection{Mitigate Latency Overheads: Dynamic Touch\'e}
One can mitigate signature-collision latency overheads by compressing only when it is useful. To this end, Touch\'e continuously monitors the average memory latency at the LLC controller. The average memory latency is defined as the total latency that is experienced by each request and this can emanate from the LLC and main memory.

\begin{figure}[h!] 
  \vspace{-0.05in}
  \centering \centerline{\psfig{file=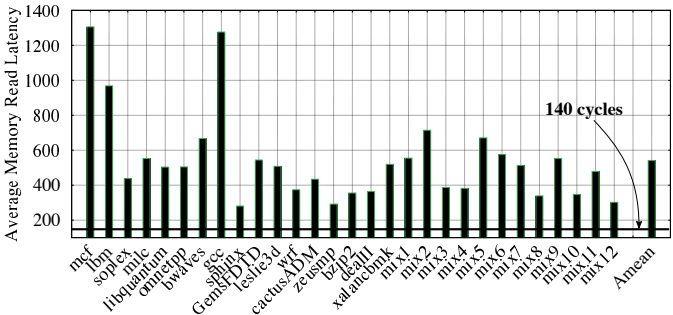,width=\columnwidth}}
      \vspace{-0.15in}
  \caption{The average memory latency for reads. On average, the average memory access latency is 541 cycles. Therefore, Touch\'e has a latency overhead of only 0.26\%.}
  \label{fig:readlat}
 \vspace{-0.2in}
\end{figure}

Touch\'e enables compression only when the average memory access latency is 100x greater than the latency overheads of signature collisions. As signature collisions increase the LLC tag access latency by 1.4 cycles, Touch\'e enables compression only when the average memory latency is greater than 140 cycles. This has two key advantages. First, Touch\'e is enabled for workloads that showcase a large memory latency and benefit from LLC compression. Second, the latency overheads from Touch\'e are capped at 1\%. As shown in Figure~\ref{fig:readlat}, for memory intensive benchmarks, the average memory latency for reads is 541 cycles (significantly higher than 140 cycles). Therefore, the latency overhead of Touch\'e is only 0.26\%.

\subsection{Superblock Marker (SMARK) Mechanism}
The SIGN Engine enables storage of up to three blocks. However, some cachelines may contain superblocks (four compressed blocks from neighboring addresses).
\subsubsection{Benefits of Including Superblocks}
\begin{figure}[h!]
  \vspace{-0.1in}
  \centering \centerline{\psfig{file=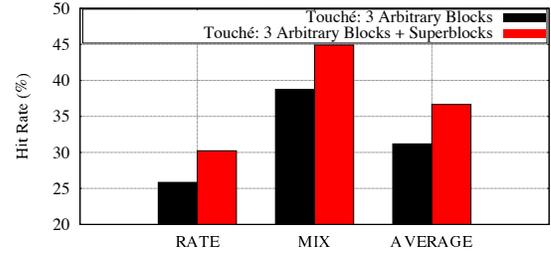,width=0.8\columnwidth}}
      \vspace{-0.1in}
  \caption{The average hit-rate of Touch\'e with 3 blocks versus 3 blocks with superblocks. On average, the hit-rate increases to 37.5\% by combining superblocks.}
  \label{fig:3to4}
  \vspace{-0.1in}
\end{figure}

Figure~\ref{fig:3to4} shows the hit-rate of Touch\'e while maintaining up to three compressed blocks and compares this against a scheme that also stores superblocks (up to four blocks). For a superblock, Touch\'e tries to compress each block to 15Bytes. This enables Touch\'e to get the benefits of storing both the superblock-tags and the arbitrary tags. If we can store superblocks and arbitrary blocks at the same time, we can increase the average hit-rate of Touch\'e from 31.5\% to 37.5\%.

\subsubsection{Identifying Potential Cachelines}
During an LLC install, if the block is compressible and if the candidate cacheline already contains compressed blocks from its neighboring addresses, then this cacheline is also a superblock candidate. The TADA mechanism is used to identify superblock candidates by extracting the full tag addresses for all the blocks in a cacheline during an LLC install.

\subsubsection{Generating Markers}
Touch\'e implements a ``Superblock Marker'' (SMARK) mechanism in the tag manager. SMARK mechanism generates a random 16-bit marker at boot-time and uses this marker throughout the operational time of the system.

\begin{figure}[h!]
  \vspace{-0.1in}
  \centering \centerline{\psfig{file=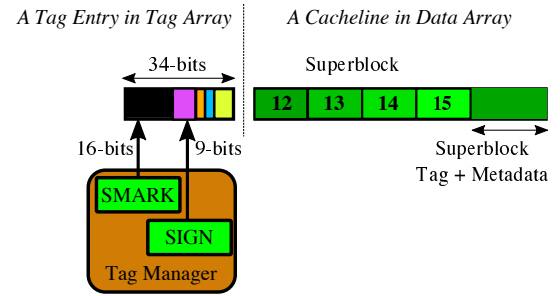,width=0.8\columnwidth}}
   \vspace{-0.1in}
  \caption{The Superblock Marker (SMARK) mechanism. The SMARK mechanism generates a unique 16-bit marker to identify superblocks. It then appends this marker with the 9-bit signature from the SIGN engine.}
  \label{fig:SMARK}
  \vspace{-0.1in}
\end{figure}

Once the TADA mechanism identifies a superblock cacheline, it informs the tag manager. The tag manager then retrieves the 16-bit marker from the SMARK mechanism. It then informs the SIGN engine to ignore the last 2-bits (corresponding to four neighboring addresses) of the full tag address to generate a unique 9-bit signature. This ensures that neighboring addresses in the superblock generate the same 9-bit signature. Thereafter, the tag manager appends the 9-bit signature to the 16-bit marker and writes the 29-bit full tag of the first block within the superblock at the end of the compressed blocks in the data array. Figure~\ref{fig:SMARK} shows the superblock-tag generation.

\begin{figure*}[t!]
      \vspace{-0.1in}
  \centering \centerline{\psfig{file=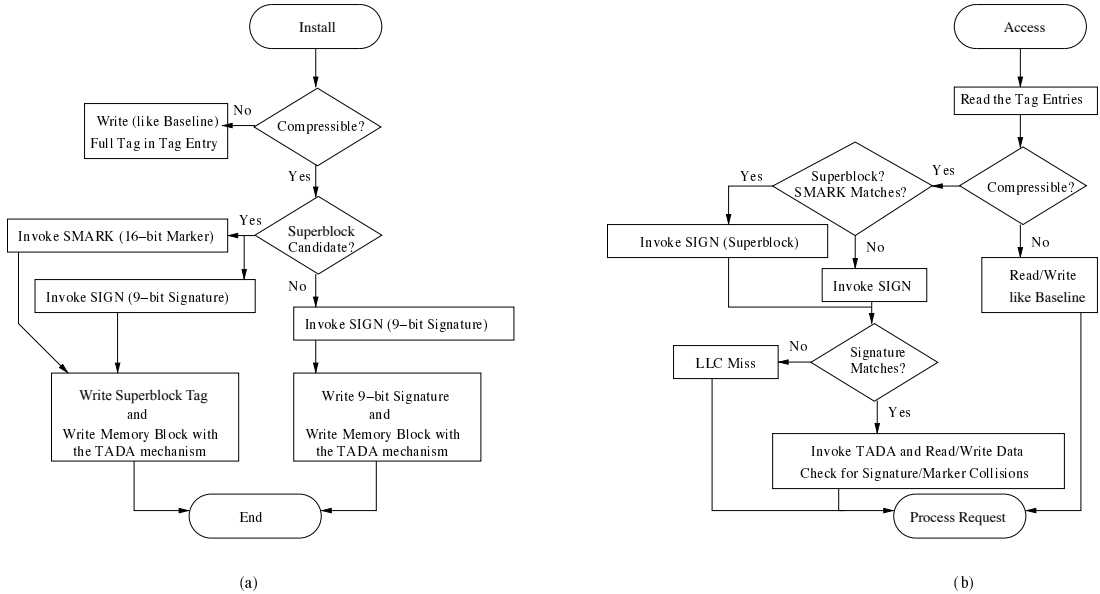,width=0.8\textwidth}}
      \vspace{-0.1in}
  \caption{The flowchart detailing the high-level operations of Touch\'e for install and access requests. (a) Shows the flowchart for install requests. (b) Shows the flowchart for access requests.}
  \label{fig:flow}
  \vspace{-0.15in}
\end{figure*}

\subsubsection{Checking for Matching Markers}
On a read, the tag manager will check for matching 16-bit marker values in all the ways that store compressed blocks within a set. If there is a marker match, then the tag manager uses the 9-bit signature (generated from by ignoring the least two significant bits) and checks for a match. 

If the signature matches, then the cacheline is read from the data array. The TADA mechanism extracts the full tag address and checks if the tag address of the LLC access is one of the superblocks. If there is a match, the block is processed by the LLC controller. It is likely, the cacheline may not contain the requested block and it may simply be a false positive match. Similar to signature collisions, we call this scenario as a marker collision.
\subsubsection{Effect of Marker Collisions}
Marker collisions are extremely rare. This is because, we use markers which are 16 bits long. For instance, in an 8-way cache, the probability of a marker collision for each access is only 0.012\% and their impact on LLC latency is negligible. Furthermore, even in the case of marker collisions, the TADA mechanism ensures that the full tag address is checked before forwarding the compressed block. Therefore, SMARK works with TADA to \emph{guarantee} correctness while storing superblocks.

\subsection{Touch\'e Operation: Reads and Writes}
Figure~\ref{fig:flow} (a) and Figure~\ref{fig:flow} (b) show the flowchart for Touch\'e for LLC accesses and install requests respectively. Touch\'e invokes the SIGN, TADA, and SMARK mechanisms only for compressed data. For uncompressible data, Touch\'e works just like the baseline. Furthermore, TADA mechanism is always invoked for LLC hits of compressed blocks. This enables Touch\'e to guarantee correctness.

\subsection{Discussion: Coherence and Replacement} In the baseline LLC, the tag entry contains metadata such as the replacement policy bits and coherence states (for private LLCs). We discuss how these affect the design of Touch\'e.

\subsubsection{Handling Cache Coherence:}~\label{touchecoherence}
Touch\'e assumes a shared LLC and therefore does not encounter coherence concerns. However, in case the LLC is private, Touch\'e would need to maintain coherence states with minimal overheads. Touch\'e stores the coherence state as well as the full tag for each compressed block in the data array. Thus, the LLC controller needs to access data array for tag matching and checking the coherence state. This operation would likely increase the latency of tag matching for the coherence request. 

However, such an operation would likely incur low performance overheads. This is because, handling most of the coherence requests tends to be \textit{off} in the critical path and the coherence state can be updated \textit{after} the critical requests are serviced. In addition, if the coherence request is the ``BusRd'' which is a read request made by another core, the current core might need to send the entire block to the requesting core anyway. In this case, the additional access to the data array does not add any overheads.

Furthermore, we can eliminate the performance impact of snooping-based coherence protocols, by simply using a directory-based coherence protocol as implemented in commercial processors~\cite{inteldir}.

\subsubsection{Handling Cache Replacement Policy:}~\label{toucherep}
Each tag entry in the baseline system is already equipped with replacement information bits. As Touch\'e stores multiple compressed blocks per cacheline, ideally, it would be preferable to equip each of these blocks with additional replacement bits in the tag entry. However, this would require us to add $3 \sim 4$ bits per compressed block in the tag entry.

To minimize the overheads for storing the replacement information, whenever a cacheline is accessed, Touch\'e only updates the original replacement bits. Touch\'e does not keep track of individual replacement bits for each block. During replacement, Touch\'e selects the victim cacheline based on the original replacement bits and randomly evicts one block from within the victim cacheline.

	\section{Experimental Methodology}
\label{sec:methodology}
\begin{figure*}[htb!]
  \vspace{-0.15in}
  \centering \centerline{\psfig{file=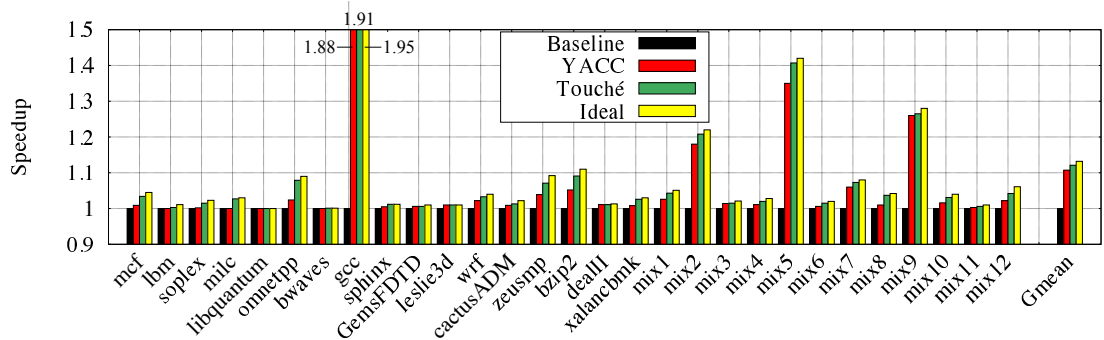,width=0.8\textwidth}}
      \vspace{-0.1in}
  \caption{Speedup of Touch\'e as compared to a baseline system that does not employ compression. On average, Touch\'e achieves a speedup of 12\% (Ideal -- 13\%, YACC -- 10.3\%) by enabling compressed blocks from arbitrary addresses to be placed next to each other while also allowing superblocks to be stored.}
  \label{fig:perf}
    
\end{figure*}

To evaluate the performance benefits of Touch\'e, we develop a trace-based simulator based on the USIMM~\cite{usimm} which is a detailed memory system simulator. We extended the USIMM to model the processor core and a detailed cache hierarchy. Our processor model supports the out-of-order (OoO) execution. Our detailed cache model supports various replacement policies such as LRU, DRRIP~\cite{Jaleel:2010:HPC:1815961.1815971}, and DIP~\cite{Qureshi:2007:AIP:1250662.1250709}. The baseline system configuration is described in Table~\ref{table:system_config}. To enable efficient compression, the compression engine modeled in the cache model employs the BDI~\cite{bdi,Pekhimenko:2013:LCP} and FPC~\cite{fpc} compression algorithms and uses the one with the best compression ratio for each cacheline. As per prior work in BDI and FPC, we assume that compression and decompression of data incurs only a single-cycle latency. We compare Touch\'e to the previous state-of-the-art scheme called YACC that uses only ``superblocks''~\cite{YACC}. We also compare our scheme against an ``Ideal'' scheme that can store either three arbitrary blocks or a superblock (four neighboring blocks) without any area overheads. The Ideal scheme uses the same replacement policy as Touch\'e (described in Section~\ref{toucherep}).
\begin {table}[thb!]
\begin{center} 
\vspace{-0.15in}
\caption{Baseline System Configuration}{
\vspace{-0.15in}
\resizebox{1\columnwidth}{!}{
\begin{tabular}{|c|c|}
\hline
                 Number of cores (OoO)         & 4     \\ 
                 Processor clock speed         & 3.2 GHz      \\
                 Issue width                   & 8         \\ 
                 \hline
                 \hline
                 L1 Cache (Private)   & 32KB, 8-Way, 64B lines, 4 cycles \\ 
                 L2 Cache (Private)   & 256KB, 8-Way, 64B lines, 12 cycles \\
                 \hline
                 \hline
                 Last Level Cache (Shared)     & 4MB, 8-Way, 64B lines \\ 
                 LLC Tag Access Latency        & 5 cycles  \\ 
                 LLC Data Access latency       & 30 cycles   \\ \hline \hline
                 Memory bus frequency              & 1600MHz (DDR 3200MHz) ~\cite{DDR4:2015}\\
                 Memory channels     & 2                    \\
                 Ranks per channel             & 1          \\
                 Banks Groups                  & 4          \\
                 Banks per Bank Group          & 4          \\
                 Rows per bank                 & 64K \\ 
                 Columns (cache lines) per row & 128       \\ \hline \hline
                 DRAM Access Timings: $T_{RCD}$-$T_{RP}$-$T_{CAS}$ & 22-22-22~\cite{lisa}\\ 
                 DRAM Refresh Timings: $T_{RFC}$ & 420ns~\cite{refpause,avatar}\\ \hline

\end{tabular}}
}
\label{table:system_config}
\end{center}
\vspace{-0.2in}
\end{table}

We chose memory intensive benchmarks, which have greater than 1 MPKI (LLC Misses Per 1000 Instructions), from the SPEC CPU2006 benchmarks. We warm up the caches for 2 Billion instructions and execute 4 Billion instructions. To ensure adequate representation of regions of compressibility~\cite{compresspoint} and performance~\cite{simpoint}, the 4 Billion instructions are collected by sampling 400 Million instructions per 1 Billion instructions over a 40 Billion instruction window. We execute all benchmarks in rate mode. We also create twelve 4-threaded mixed workloads by forming two categories of SPEC2006 Benchmarks, low MPKI, and high MPKI. As described in Table~\ref{table:mix_worload}, we randomly pick one benchmark from each category to form high MPKI mixed workloads and medium MPKI mixed workloads. We perform timing simulation until all the benchmarks in the workload finish execution.

\begin {table}[h!]
\begin{center} 
\caption{Workload Mixes}{
\vspace{-0.15in}
\resizebox{0.7\columnwidth}{!}{
\begin{tabular}{|c||c|}
\hline
mix1 & mcf, libquantum, GemsFDTD, wrf \\
mix2 & lbm, gcc, bzip2, bwaves\\
mix3 & milc, sphinx, leslie3d, zeusmp\\
mix4 & soplex, omnetpp, cactusADM, dealII\\
mix5 & xalancbmk, mcf, gcc, sphinx\\
mix6 & omnetpp, lbm, milc, xalancbmk\\
mix7 & astar, mcf, milc, calculix\\
mix8 & omnetpp, gobmk, sjeng, libquantum\\
mix9 & namd, gcc, lbm, dealII\\
mix10 & soplex, tonto, hmmer, perlbench\\
mix11 & GemsFDTD, bwaves, povray, zeusmp\\
mix12 & wrf, xalancbmk, h264, gamess\\
\hline
\end{tabular}}
}
\label{table:mix_worload}
\end{center}
\end{table}

	
\begin{figure*}[htb!]
  \centering \centerline{\psfig{file=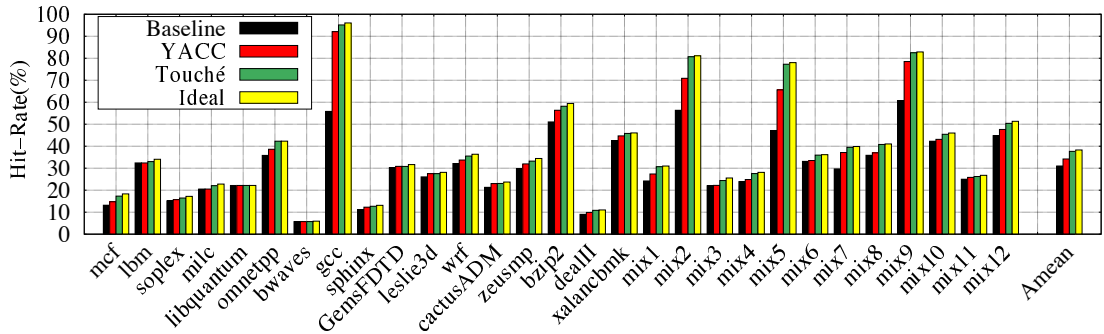,width=0.8\textwidth}}
      \vspace{-0.1in}
  \caption{The Hit-Rate of Touch\'e as compared to a baseline system that does not employ compression. On average, Touch\'e increases the hit-rate by 6\% points (Ideal -- 7\% points, YACC -- 4\% points) by storing a larger number of blocks within the LLC.}
  \label{fig:hitrate}
\end{figure*}

\begin{figure*}[t!]
  \vspace{-0.1in}
  \centering \centerline{\psfig{file=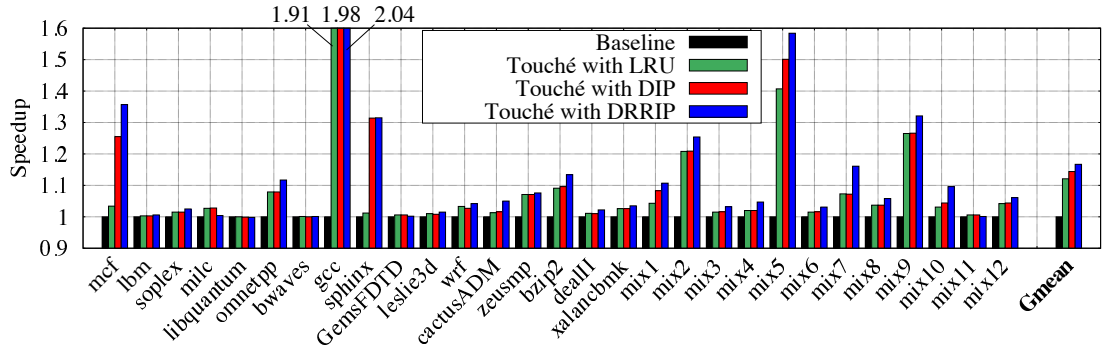,width=0.8\textwidth}}
  \vspace{-0.1in}
  \caption{The sensitivity of Touch\'e to different replacement policies. As Touch\'e is only a LLC compression framework, it is orthogonal to the replacement policy. Touch\'e shows an increasing speedup of 12\%, 14.5\%, and 16.7\% for the LRU, DIP, and DRRIP replacement policies respectively.}
  \label{fig:repl}
  \vspace{-0.1in}
\end{figure*}

\section{Results}
This section discusses the performance, hit-rate, and sensitivity results of Touch\'e.
\subsection{Performance Impact}
Figure~\ref{fig:perf} shows the speedup of Touch\'e when compared to a baseline system that does not employ compression. On average, Touch\'e has a speedup of 12\%. Ideally, when we can place compressed memory blocks without any area overheads in tag and data arrays, we get a speedup of 13\%. On the other hand, YACC achieves 10.3\% speedup by capturing the superblocks. Our analysis shows that \textit{gcc} benefits the most from LLC compression. \textit{gcc} is extremely sensitive to the LLC capacity and as the miss rate of \textit{gcc} drops from 45\% to 5\% (by 9x) due to Touch\'e, \textit{gcc} experiences very low memory access latency. This is because, at 5\% miss-rate, almost all of its working set now fits in the LLC. Therefore, \textit{gcc} shows a speedup of 91\% due to Touch\'e. For all other workloads, the drop in \textit{miss-rate} is at most 2.4x (see Figure~\ref{fig:hitrate}), hence they show up to 50\% speedup.

\subsection{Effect on Last-Level Cache Hit-Rate}
Figure~\ref{fig:hitrate} shows the speedup of Touch\'e when compared to a baseline system that does not employ compression. On average, Touch\'e increases the hit rate by 6\% points. In the ideal case, when we can place compressed memory blocks from arbitrary addresses without any area overheads in tag and data arrays, the hit-rate increases 7\% points. On the other hand, YACC increases the hit-rate by 4\% points. Furthermore, some workloads like \textit{gcc}, \textit{mix2}, \textit{mix5}, and \textit{mix9} get significant increase in hit rate. 

We also observe that hit-rates either increase or remain the same for benchmarks. Furthermore, Touch\'e closely follows the hit-rate of an ideal LLC compression technique. The slight loss in hit-rate from the ideal LLC compression is due to the capacity loss in the data array from the TADA mechanism.

\subsection{Sensitivity to Replacement Policy}
As Touch\'e is a LLC compression technique, it does not interfere with the replacement policy. Typically, the LLC controller will choose a cacheline based on its replacement policy. Touch\'e then evicts a block from within the selected cacheline randomly. Therefore, replacement policies are orthogonal to the Touch\'e framework.

Figure~\ref{fig:repl} shows the speedup of Touch\'e for different cache replacement policies. On average, Touch\'e increases the speedup from 12\% while using LRU, to 14.5\% while using DIP. The speedup is increased to 16.7\% while using DRRIP replacement policy. Therefore, irrespective of the replacement policy, Touch\'e continues to provide high performance by enabling efficient compression.

\subsection{Impact on Memory Latency}
\begin{figure}[h!]
   \vspace{-0.05in}
  \centering \centerline{\psfig{file=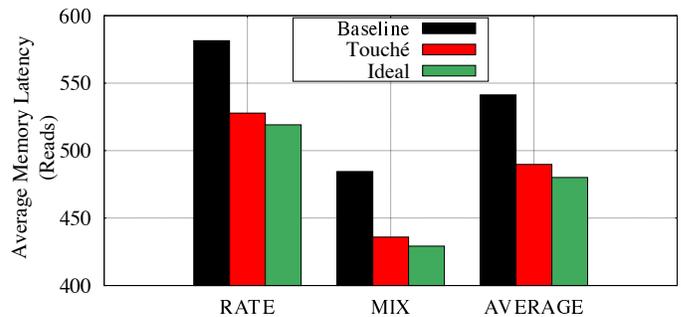,width=\columnwidth}}
    \vspace{-0.15in}
  \caption{The average memory read latency for Touch\'e. On average, Touch\'e reduces the memory read latency from 541 cycles to 489 cycles.}
  \label{fig:MemLat}
  \vspace{-0.2in}
\end{figure}

Figure~\ref{fig:MemLat} shows the impact of Touch\'e on the average memory latency for reads. As Touch\'e provides a higher LLC hit rate, it also reduces the average memory read latency. On average, Touch\'e reduces the memory read latency from 541 cycles to 489 cycles. In the ideal case, we can reduce the average memory read latency to 478 cycles as this scheme provides slightly higher hit-rate .




\subsection{Sensitivity to Last-Level Cache Size}
\begin{figure}[h!]
      \vspace{-0.1in}
  \centering \centerline{\psfig{file=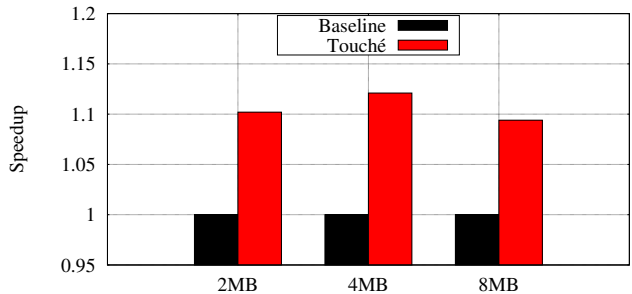,width=\columnwidth}}
      \vspace{-0.15in}
  \caption{The sensitivity of Touch\'e to the size of the LLC. Even after varying the LLC size, Touch\'e continues to provide at least 9\% average speedup.}
  \label{fig:LLCSize}
    \vspace{-0.15in}
\end{figure}

Figure~\ref{fig:LLCSize} shows the impact of LLC size on the effectiveness of Touch\'e. Touch\'e is robust to different LLC sizes and continues to be effective. For instance, even while using a 2MB cache, Touch\'e provides an average speedup of 10\%. Even after doubling the LLC size from 4MB to 8MB, Touch\'e still provides a 9\% average speedup. 

\subsection{Impact on Low-MPKI Benchmarks}
\begin{figure}[h!]
    \vspace{-0.1in}
  \centering \centerline{\psfig{file=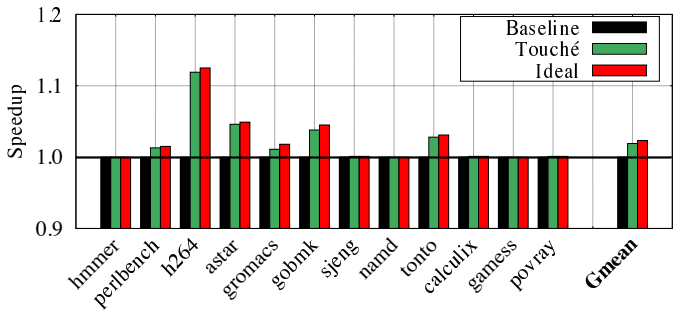,width=\columnwidth}}
      \vspace{-0.1in}
  \caption{Impact of Touch\'e on low MPKI workloads. Touch\'e does not hurt the performance of any low MPKI workload. Touch\'e provides an average speedup of 1.9\% for these workloads.}
  \label{fig:LMPKI}
\end{figure}

Until now, we have presented results only for high MPKI benchmarks. However, for implementation purposes, it is vital that Touch\'e does not hurt the performance of low MPKI benchmarks. Figure~\ref{fig:LMPKI} shows the impact of Touch\'e on the performance of Low MPKI workloads from the SPEC2006 suite. Overall, Touch\'e does not cause slowdown for any Low MPKI workload. On the contrary, Touch\'e provides an average speedup of 1.9\% for these workloads.


\section{Related Work} 
In this section, we describe prior work that is closely related to the ideas discussed in this paper.

\subsection{Efficient Compression Algorithms}
Cache compression algorithms like Frequent Pattern Compression (FPC)~\cite{fpc}, Base-Delta-Immediate (BDI)~\cite{bdi}, and Cache Packer (C-PACK)~\cite{cpack} have low decompression latency and require low implementation cost (i.e., area overhead). The C-PACK algorithm can be improved further by detecting zero cache lines~\cite{SCC}. Recently, Kim et al.~\cite{7551404} introduce a bit-plane compression algorithm that uses a bit-plane transformation to achieve a high compression ratio. Touch\'e is orthogonal to all of these compression algorithms. Touch\'e can select any of these algorithms to meet the hardware budget, latency constraints, and application's requirements.

\subsection{Cache Compression with Tag Management}
Prior works have proposed compressed cache architectures to improve the effective cache capacity~\cite{Alameldeen:2004,SCC,YACC,DCC1,DCC2}. For instance, a variable-size compressed cache architecture using FPC was proposed ~\cite{Alameldeen:2004}. This architecture doubles the cache size when all cachelines are compressed while requiring twice as many tag entries.

To reduce tag overhead of the compressed cache, DCC~\cite{DCC1} and SCC~\cite{SCC} use superblocks to track multiple neighbor blocks with a single tag entry. Recently, YACC~\cite{YACC} was proposed to reduce the complexity of SCC by exploiting the compression and spatial locality. YACC still restricts the mapping of compressed cachelines as it requires superblocks that contain cachelines only from neighboring addresses. Furthermore, YACC requires that those cachelines be of the same compressed size. Touch\'e eliminates this fundamental limitation of the super block-based compressed cache. On average, YACC provides 10.3\% speedup while requiring additional bits in the tag area resulting in 1.35x tag area. Touch\'e provides 12\% speedup without any area overheads. To increase LLC efficiency, Amoeba-Cache~\cite{ameoba}, proposes storing tag and data together while eliminating the tag area. However, to create space for tags, Amoeba-Cache stores only parts of the memory block within the cache. As DRAM caches do not encounter tag storage problems and tend to be bandwidth sensitive, Young et. al.~\cite{Young:2017} use compression in DRAM caches to improve both capacity and bandwidth dynamically.

\subsection{Compression using Deduplication}
Data deduplication exploits the observation that several memory blocks in the LLC contain the same identical value~\cite{Dedup,DDE,DupManage}. To improve efficiency, these techniques store only a single value of these memory blocks within the LLC and design techniques to maintain tags that point to such memory blocks. 

Exploit the presence of identical memory blocks in the LLC, Dedup~\cite{Dedup} changes the LLC to enable several tags to point to the same data. To this end, the tag array is decoupled from the data array. Each tag entry is then equipped with pointers to enable them to point to arbitrary memory blocks in the data array. Touch\'e is orthogonal to Dedup, as Touch\'e is compression technique that compresses arbitrary memory blocks independently and enables Dedup to be applied over it.

\subsection{Main Memory Compression Techniques}
Compression can also be used for main memory. Memzip compresses data for improving the bandwidth of the main memory~\cite{memzip}. Pekhimenko et. al.~\cite{Pekhimenko:2013:LCP} and Abali et. al~\cite{abali} have proposed efficient techniques to improve the effective capacity of main memory using compression. Compression can also be used in Non-Volatile Memories (NVM) to reduce energy and improve performance~\cite{7446056}. As compression increases the number of bit-toggles on the bus, Pekhimenko et. al.~\cite{7446064} minimizes bit-toggles and reduces the bus energy consumption. Recently, Compresso~\cite{compresso} memory system was proposed to reduce the additional data movement caused by metadata accesses for additional translation, changes in compressibility of the cacheline, and compression across cacheline boundaries. Similarly, DMC~\cite{DualMem} was proposed to improve memory capacity.

Compression can use software support and increase the main memory capacity. Products like IBM MXT and VMWare ESX use ``Balloon Drivers" to allocate and hold unused memory when data becomes incompressible or when Virtual Machines exceed capacity thresholds~\cite{IBMRC,Franaszek-IBM,mxt,pinnacle,vmwareunderstanding}. One can also use compression in the context of memory security. Morphable Counters~\cite{morphcntr} compress integrity tree and encryption counters to reduce the size and height of the integrity tree within the main memory.

\subsection{Metadata Management for Main Memory}
To reduce metadata bandwidth overheads from compression, Attach\'e~\cite{attache} and PTMC~\cite{PTMC} enables data and metadata to be accessed together. Deb et. al.~\cite{aliecc} describes the challenges in maintaining metadata in main memory and recommend using ECC to store Metadata. While this technique is useful for memory modules that have ECC in them, LLC uses tag entries and does not have to rely on ECC to store metadata~\cite{Yoon:2011:AGM,archshield,faultsim1,faultsim2,process-sig,xed,morphecc,citadel1,citadel2,synergy}. However, Touch\'e can be expanded to include the ECC within LLC to store metadata.

\subsection{Other Relevant Work}
Sardashti and Wood~\cite{Sardashti:2017:CGU:3154814.3138805} observed that cachelines in the same page may not have similar compressibility. 
Hallnor et. al.~\cite{stevencomp} proposed using compressed data throughout the memory hierarchy. This approach reduces the overheads of compression and decompression at every level of memory hierarchy. 
Sathish et al.~\cite{satgpu} try to save memory bandwidth by using both lossy and lossless compression for GPUs. Recent work from Han et. al.~\cite{DBLP:journals/corr/HanMD15} and Kadetotad et. al.~\cite{Kadetotad} used compression with deep neural networks to significantly improve performance and reduce energy. These prior work are orthogonal to Touch\'e.

Cache compression has also been used to reduce cache power consumption. Residue cache architecture~\cite{ResidueCache} reduces the last-level cache area by half, resulting in power saving. Other prior works have been proposed to lower the negative impacts of compression on the replacement. ECM~\cite{ECM} reduces the cache misses using Size-Aware Insertion and Size-Aware Replacement. CAMP~\cite{pekhimenko2015} exploits the compressed cache block size as a reuse indicator. Base-Victim~\cite{bvc} was also proposed to avoid performance degradation due to compression on the replacement. The power-performance efficiency of Touch\'e can be improved using these prior work.

	\section{Summary}
The Last-Level Cache (LLC) capacity per core has stagnated over the past decade. One way to increase the effective capacity of LLC is by employing data compression. Data compression enables the LLC controller to pack more memory blocks within the LLC. Unfortunately, the additional compressed memory blocks require additional tag entries. The LLC designer needs to provision additional tag area to store the tag entries of compressed blocks. We can also restrict data placement within each cacheline to neighboring addresses (superblocks) and reduce the tag area overheads. Ideally, we would like to get the benefits of LLC compression without incurring any tag area overheads.

To this end, this paper proposes Touch\'e, a framework that enables LLC compression without any area overheads in the tag or data arrays. Touch\'e uses shortened signatures to represent full tag address and appends the full tags to the compressed memory blocks in the data array. This enables Touch\'e to store arbitrary memory blocks as neighbors. Furthermore, Touch\'e can be enhanced further to include superblocks. Touch\'e is completely hardware based and achieves a near-ideal speedup of 12\% (ideal 13\%) without any changes or area overheads to the tag and data array.

\section*{Acknowledgement}
We thank the anonymous reviewers for their feedback. We thank Amin Azar his feedback on compression. 
This work was partially supported by the Natural Sciences and Engineering Research Council of Canada (NSERC) [funding reference number RGPIN-2019-05059] and by the National Research Foundation of Korea (NRF) grant funded by the Korea government (MSIT) [funding reference number NRF-2019R1G1A1011403].

	\balance
	\bibliographystyle{IEEEtran}
	\bibliography{ref}

\end{document}